\documentclass[english]{IEEEtran}
\usepackage[T1]{fontenc}
\usepackage[latin9]{inputenc}
\usepackage{color}
\usepackage{framed}
\usepackage{textcomp}
\usepackage{amsthm}
\usepackage{amsmath}
\usepackage{graphicx}
\definecolor{shadecolor}{rgb}{0.75,0.75,0.75}
\usepackage{amssymb}
\usepackage{esint}

\makeatletter

\DeclareRobustCommand{\greektext}{%
  \fontencoding{LGR}\selectfont\def\encodingdefault{LGR}}
\DeclareRobustCommand{\textgreek}[1]{\leavevmode{\greektext #1}}
\DeclareFontEncoding{LGR}{}{}
\DeclareTextSymbol{\~}{LGR}{126}
\newcommand{\lyxmathsym}[1]{\ifmmode\begingroup\def\b@ld{bold}
  \text{\ifx\math@version\b@ld\bfseries\fi#1}\endgroup\else#1\fi}

\theoremstyle{plain}
\newtheorem{thm}{Theorem}
\theoremstyle{definition}
\newtheorem{defn}[thm]{Definition}
\theoremstyle{plain}
\newtheorem{lem}[thm]{Lemma}

\usepackage{dsfont}
\usepackage{mathrsfs}
\usepackage{epsfig}

\makeatother

\usepackage{babel}

\begin{document}

\title{Fair Scheduling in Networks Through Packet Election}

\author{Srikanth Jagabathula and Devavrat Shah%
\thanks{This work was supported in parts by NSF CAREER CNS 0546590 and NSF
CCF 0728554.%
}%
\thanks{The authors are with the Department of Electrical Engineering and
Computer Science, Massachusetts Institute of Technology, Cambridge,
MA, 02139 USA e-mail: \{jskanth, devavrat\}@ mit.edu%
}}
\maketitle
\begin{abstract}
We consider the problem of designing a fair schedul-ing algorithm
for discrete-time constrained queuing networks. Each queue has dedicated
exogenous packet arrivals. There are constraints on which queues can
be served simultaneously. This model effectively describes important
special instances like network switches, interference in wireless
networks, bandwidth sharing for congestion control and traffic scheduling
in road roundabouts. Fair scheduling is required because it provides
isolation to different traffic flows; isolation makes the system more
robust and enables providing quality of service. Existing work on
fairness for constrained networks concentrates on flow based fairness.
As a main result, we describe a notion of packet based fairness by
establishing an analogy with the ranked election problem: packets
are voters, schedules are candidates and each packet ranks the schedules
based on its priorities. We then obtain a scheduling algorithm that
achieves the described notion of fairness by drawing upon the seminal
work of Goodman and Markowitz (1952). This yields the familiar Maximum
Weight (MW) style algorithm. As another important result, we prove
that the algorithm obtained is throughput optimal. There is no reason
a priori why this should be true, and the proof requires non-traditional
methods\end{abstract}
\renewcommand{\leftmark}{IEEE TRANSACTIONS ON INFORMATION THEORY} 

$ $\global\long\def\crS{\mathscr{S}}

\global\long\def\bbone{\mathds{1}}

\global\long\def\dotprod#1#2{\left\langle #1,#2\right\rangle }

\global\long\def\abs#1{\left|#1\right|}

\global\long\def\E#1{\mathbb{E}\left[#1\right]}

\global\long\def\relint{\textrm{\ensuremath{\mathop{\mathrm{relint}}}}}

\global\long\def\co#1{\mathop{\mathrm{co}\left(#1\right)}}

\global\long\def\v#1{\mathsf{value}(#1)}

\global\long\def\ind#1{\bbone_{\left\{  #1\right\}  }}

\section{Introduction}

In this paper, we focus on the problem of scheduling in constrained
queuing networks. Specifically, we consider a collection of queues
operating in discrete time with constraints on which queues may be
served simultaneously. Such queuing systems serve as effective modeling
tools for a large array of important practical problems, and their
performance is crucially dependent on the effectiveness of the scheduling
algorithm.

In this setup, the basic question is to design a scheduling algorithm
that is optimal. There are several performance criteria, with often
inherent trade-offs, that determine the optimality of a scheduling
algorithm. The first is throughput optimality. A queuing system has
a limited amount of resources. The natural constraints imposed result
in an inherent limitation on the amount of traffic load that can be
supported. This is called the capacity of the system. Roughly speaking,
a scheduling algorithm that achieves the capacity utilizes system
resources optimally. Such an algorithm is called throughput optimal.

Apart from being throughput optimal, a scheduling algorithm should
allocate resources in a fair manner. The queuing system is a common
resource shared by various traffic flows, and the scheduling algorithm
should ensure that no flow is receiving more than its \textquotedblleft{}fair\textquotedblright{}
share of resources. It is important to realize that fairness in queuing
systems is not only an intuitively desired goal but also one with
an immense practical impact.

A very important consequence of fairness is isolation of various traffic
flows. Throughput optimality is oblivious to the identities of the
different flows. But, identities are important for the following two
important reasons: (1) Being oblivious to flow identities, throughput
optimal algorithms often favor flows with a high data rate. Therefore,
a particular flow might ill-behave and flood the system with a high
data rate maliciously resulting in the deterioration of service to
other flows. Since the system is a shared resource, the algorithm
should identify the rogue flow and limit the negative impact on well-behaved
flows. (2) Secondly, isolation is important to provide performance
guarantees, and thereby Quality of Service (see Keshav (1997) \cite{Keshavbook97}),
to various flows in the system. Designing a scheduling algorithm that
is fair will overcome these issues. Other benefits of fairness include
reducing burstiness of flows, eliminating bottlenecks and reducing
the impact of certain kinds of Denial-of-Service (DoS) attacks (see
Bonald and Massoulie\textasciiacute{} (2001) \cite{BM01}, and Yau
et al (2005) \cite{YLLY05}). In essence, a fair scheduling algorithm
makes the queuing system robust and less prone to manipulation by
individuals.

A natural way to achieve isolation among flows, in order to provide
protection and performance guarantees, is to dedicate resources to
each of the flows. In fact, this is the approach taken in most of
the work done on designing fair algorithms for input queued switches
(details in Section~\ref{sub:Related-work}). This approach, though,
is limited for the following reasons: Firstly, because of constraints,
it is not straightforward to determine the amount of resources to
be allocated to each flow in a queuing network. Moreover, such determination
would require the knowledge of flow arrival rates; whereas, in the
spirit of being implementable, we require the scheduling algorithm
to be online i.e., use only current network state information like
queue-sizes, age of packets, etc., and myopic i.e., oblivious to flow
arrival rates. Secondly, resource allocation takes place on an average
over a long period of time. This is appropriate in a flow level model
where the arrival statistics remain constant for long time periods.
This assumption, though, is questionable in many applications like
Internet traffic where short flows predominate. 

We note that designing a fair scheduling algorithm comprises two sub-problems:
defining a reasonable notion of fairness, and designing an algorithm
to achieve the desired notion of fairness. The notion of fairness
is utilized to determine the resources (more specifically, the rate
or bandwidth) to be allocated, and the scheduling algorithm ensures
that the resources are allocated on an average over a long period
of time. Inspired by this, our approach would be to first define a
notion of packet based fairness and then design a scheduling algorithm
to achieve the defined notion of fairness. For obvious reasons, we
also need to reconcile the benefits of fairness with achieving throughput
optimality.

Motivated by the above discussion, we attempt to achieve the following
three goals in this paper: (1) Define a notion of packet based fairness.
(2) Design an online and myopic algorithm, that is also throughput
optimal, to achieve the notion of fairness. (3) Provide the throughput
optimality proof of the algorithm.

\subsection{Our contributions}

The need for a packet based notion of fairness that can be used in
a constrained network is clear. But, defining a precise mathematical
notion of fairness that achieves the desired intuitive and practical
benefits of fairness is quite challenging. Unfortunately, none of
the existing notions of fairness directly extend to a packet based
constrained queuing network. Existing notions of flow based fairness
are based on the utility maximization framework (proposed by Kelly,
Maullo and Tan (1998) \cite{KMT98}), which is a concept borrowed
from Economics literature. In a similar spirit, we define a notion
of fairness by establishing a novel analogy between scheduling in
constrained queuing networks and a ranked election problem%
\footnote{A ranked election problem deals with choosing a winning permutation
of candidates using a set of votes, where each vote is a permutation
of the candidates. Refer to Section~\ref{sec:Ranked-Election}.%
}. Ranked election is a widely studied problem in the Economics (and
Political Science) literature and this analogy provides a ready framework
to leverage this work. We draw upon the work of Goodman and Markowitz
(1952)~\cite{GM52} to obtain a unique characterization of the schedule.
This, rather surprisingly, yields a maximum weight (MW) style algorithm.
MW style algorithms are very popular in the literature and are very
well understood. Thus, MW algorithms choose schedules in a \textquotedblleft{}fair\textquotedblright{}
manner, though the definition of \textquotedblleft{}fair\textquotedblright{}
for different weights is different. It should be noted that the choice
of weights is crucial for obtaining the intuitive desirable properties
of fairness, and we make an important contribution here.

As another important contribution, we prove that our algorithm is
throughput optimal. There is no a priori reason for this to be true.
Even though the algorithm we design is the familiar MW style algorithm,
it is not queue size or waiting time based. Therefore traditional
methods of proving throughput optimality, which include the popular
Lyapunov-Foster method and Fluid models, cannot be applied in a direct
manner. The proof technique we introduce to prove throughput optimality
is potentially applicable to a wider class of problems.

\subsection{Related work\label{sub:Related-work}}

We first begin with the work on single queue fairness. Fair scheduling
in single queues has been widely studied since the early 1990s. In
one of the earliest works, John Nagle (1987) \cite{N87} proposed
a fair algorithm for single queues called \textquotedblleft{}Fair
Queuing.\textquotedblright{} As mentioned earlier, fair scheduling
is required to minimize starvation and limit the negative impact of
rogue sources. In order to achieve this, Nagle proposed maintaining
separate queues for different flows and serving them in a round-robin
fashion. This is a great and simple to implement solution, but it
works only when all the packets are of equal size. In order to overcome
this problem Demers, Keshav and Shenker (1989) \cite{DKS89} proposed
the notion of Weighted Fair Queueing (WFQ) and its packetized implementation.
Parekh and Gallager (1993, 1994) \cite{PG93,PG94} analyzed the performance
of this packetized implementation and showed it to be a good approximation
of Generalized Processor Sharing (GPS). Shreedhar and Varghese (1996)
\cite{SV96} designed a computationally efficient version of weighted
fair queuing called Deficit Weighted Round Robin (DWRR). Even though
all these algorithms are fair, they are very complex and expensive
to implement. Hence, there was a lot of work done on achieving approximate
fairness for Internet routers through FIFO queuing and appropriate
packet dropping mechanisms. Examples include RED by Floyd and Jacobson
(1993) \cite{FJ93}, CHoKe by Pan, Prabhakar and Psounis (2000) \cite{PPP00},
and AFD by Pan, Prabhakar, Breslau and Shenker (2003) \cite{PPBS03}.

To address the issue of fairness in a network, Kelly, Maullo and Tan
(1998) \cite{KMT98} proposed a flow-level model for the Internet.
Under this model, the resource allocation that maximizes the global
network utility provides a notion of fair rate allocation. We refer
an interested reader to survey-style papers by Low (2003) \cite{L03}
and Chuang et.al. (2006) \cite{CLCD06} and the book by Srikant (2004)
\cite{S-book} for further details. We take a note of desirable throughput
property of the dynamic flow-level resource allocation model (see
for example, Bonald and Massoulie\textasciiacute{} (2001) \cite{BM01},
and de Veciana, Konstantopoulos and Lee (2001) \cite{DKL01}). This
approach, though valid for a general network with arbitrary topology,
does not take scheduling constraints into account.

We next review the work done on the design of fair scheduling algorithms
for Input Queued (IQ) switches. Switches are the most simple \textendash{}
at the same time, highly non-trivial \textendash{} examples of constrained
networks. They form the core of Internet routers and there is extensive
literature dealing with the design and analysis of various switch
architectures and scheduling algorithms. A switch is essentially a
bipartite network with input ports and output ports. The function
of a network switch is to move packets from the input ports to the
output ports using the switch fabric, just like the traffic at a traffic
junction. Depending on the placement of buffers and the switch fabric,
there are mainly two kinds of switch architectures \textendash{} input
queued switches (IQ) and output queued (OQ) switches. As their names
suggest, input queued switches have buffers only at input ports, while
output queued switches have buffers only at the output ports. The
input queued switch has a cross-bar switch fabric that imposes the
following natural constraints: only one packet can be moved from (to)
each input (output) port in each time slot. On the other hand, since
an output queued switch has buffers only at output ports, packets
arriving at the input ports are immediately transferred to their respective
output buffers. Thus, there are no scheduling constraints at the switch
fabric in an output queued switch. Because of this, the memory in
an output queued switch has to operate much faster than the memory
in an input queued switch. In most high-speed switches, memory bandwidth
is the bottleneck and hence input queued switches are more popular
and widely deployed, while output queued switches are idealized versions
that are easy to study.

It is clear from the description that scheduling in output queued
switches is equivalent to that of single queues. Hence, fair scheduling
in output queued switches just corresponds to implementing single
queue fair algorithms at different output queues. Unfortunately, such
extension is not possible for input queued switches because of the
presence of constraints. One approach is to emulate the performance
of an OQ switch by means of a IQ switch running with a minimal speedup.
An IQ switch is said to be running with a speedup $S$ if it can be
scheduled $S$ times in each time slot. This approach was taken by
Prabhakar and McKeown (1999) \cite{PM99} and Chuang, Goel, McKeown
and Prabhakar (1999) \cite{CGMP99}, where they showed that essentially
a speedup of 2 is necessary and sufficient for emulating an OQ switch.
With the OQ switch operating under any of the various policies like
FIFO, WFQ, DWRR, strict priority, etc. fairness can be achieved. Equivalently,
if an IQ switch is loaded up to 50\% of its capacity and the notion
of fairness is defined by policies like FIFO, WFQ, DWRR, strict priority,
etc., then by emulating an OQ switch with these policies, it is possible
to have fair scheduling for the IQ switch. However, for higher loading
this approach will fail due to inability of emulating an OQ switch.

This necessitates the need for defining an appropriate notion of fairness
that cleverly, and in a reasonable manner, combines the preferences
of packets based on some absolute notions along with the scheduling
constraints. In principle, this question is very similar to the question
answered by utility maximization based framework for bandwidth allocation
in a flow network. In fact, most of the existing literature on fair
scheduling algorithms for input-queued switches is concerned with
the notion of flow-based fairness. In these approaches, a flow is
identified with all the packets corresponding to an input-output pair.
There are two main approaches taken in the literature for the design
of fair algorithms for IQ switches. One class of fair algorithms implement
a fair scheduling scheme at each of the servers in the switch and
then carry out an iterative matching. This approach is based on the
Distributed Packet Fair Queuing architecture. Examples of this approach
include iPDRR proposed by Zhang and Bhuyan (2003) \cite{ZB03}, MFIQ
proposed by Li, Chen and Ansari (1998) \cite{LCA98}, and iFS proposed
by Ni and Bhuyan (2002) \cite{NB02}. This approach completely ignores
fairness issues that arise because of scheduling constraints and hence
need not guarantee an overall fair bandwidth allocation. In order
to overcome this, Hosaagrahara and Sethu (2005) \cite{HS05} and more
recently Pan and Yang (2007) \cite{PY07} propose algorithms to calculate
overall max-min rates of different flows, taking into account contention
at all levels. But this approach requires knowledge of rates of flows
and hence, the system should either learn these rates or know them
a priori.

Thus, most of the literature on the design of fair scheduling algorithms
for constrained networks is limited because it either ignores fairness
issues caused due to scheduling constraints or directly borrows flow-based
fairness notions and allocates bandwidth accordingly. Here, it is
important to emphasize the limitations of a flow-based approach: (a)
network traffic predominantly contains \textquotedblleft{}short-flows\textquotedblright{},
while flow-based approach requires existence of ever-lasting traffic
thereby inducing huge delays when applied naively, (b) flow-based
approach requires knowledge of traffic rates, which it may have to
learn, (c) our unit of data is a packet and modeling it as a flow
is just an approximation, and (d) packets have priorities and they
lack explicit utility functions.

In summary, our question is inherently combinatorial which requires
dealing with hard combinatorial constraints unlike the resource allocation
in a flow network which deals with soft capacity constraints in a
continuous optimization setup.

\subsection{Organization}

The rest of the paper is organized as follows: Section~\ref{sec:Model-and-Notation}
describes the model, introduces the notation and states the problem
formally. Section~\ref{sec:Our-Approach} motivates and describes
our approach. Section~\ref{sec:Ranked-Election} takes a digression
into Economics literature to explain the ranked election problem.
Section~\ref{sec:Analogy-between-fair} establishes the analogy between
the ranked election problem and network scheduling. Sections~\ref{sec:Most-Urgent-Cell}
and \ref{sec:Throughput-of-MUCF} present the main results of this
paper. Section~\ref{sec:Most-Urgent-Cell} formally states our algorithm,
while Section~\ref{sec:Throughput-of-MUCF} provides the details
of the proof of throughput optimality. We provide some simulation
results in Section~\ref{sec:Experiments} and then finally conclude
in Section~\ref{sec:Conclusion}.

\section{Model and Notation\label{sec:Model-and-Notation}}

We now describe a generic abstract model of a constrained queuing
network. The model corresponds to a single-hop network. This generic
model describes important special instances like an input queued switch,
wireless network limited by interference, congestion control in TCP
or even traffic in a road junction. In each of these instances, the
model effectively captures the constraints imposed by nature on simultaneous
servicing of queues. We will describe the examples of an input queued
switch and a wireless network in detail. We focus on these two examples
because they encapsulate a large class of scheduling problems.

\subsection{Abstract formulation}

Consider a collection of $N$ queues. Time is discrete and is indexed
by $\lyxmathsym{\textgreek{t}}\in\left\{ 0,1,\ldots\right\} $. Each
queue has a dedicated exogenous process of packet arrival. The arrival
processes of different queues are independent. All packets are assumed
to be normalized to unit length. Arrivals to each queue occur according
to a Bernoulli process.

The service to the queues is subject to scheduling constraints in
that not all queues can be served simultaneously. The scheduling constraints
present are described by a finite set of feasible schedules $\crS\subseteq\left\{ 0,1\right\} ^{N}$
. In each time slot a feasible schedule $\lyxmathsym{\textgreek{p}}\in\crS$
is chosen and queue $n$ is offered a service $\lyxmathsym{\textgreek{p}}_{n}$
in that time slot. Since each packet is of unit length, when a non-empty
queue receives service, a packet departs from the queue. We assume
that $\crS$ is monotone i.e., if $\lyxmathsym{\textgreek{p}}\in\crS$,
then for any $\lyxmathsym{\textgreek{sv}}\leq\lyxmathsym{\textgreek{p}}$
component-wise i.e., $\lyxmathsym{\textgreek{sv}}_{n}\leq\lyxmathsym{\textgreek{p}}_{n}$,
$\lyxmathsym{\textgreek{sv}}\in\crS$. Further, we assume that for
each $n$, there exists a schedule $\pi\in\crS$ that serves it i.e.,
$\pi_{n}=1$.

\begin{figure}
\includegraphics[scale=0.5]{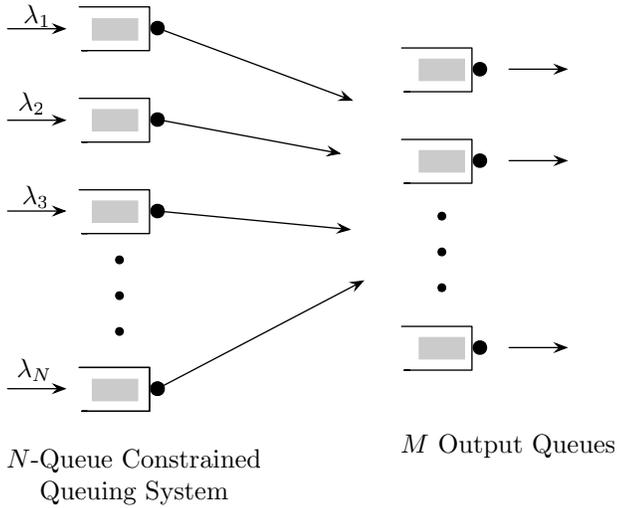}

\caption{Abstract model of the constrained queuing system \label{fig:model}}

\end{figure}

Packets exit the system from any of the $M$ output lines. The lines
are assumed to operate at unit speed and hence at most one packet
can leave the network from each line in each time slot. Each of the
$M$ output lines maintain buffers termed output queues to store packets
that are ready to depart. We assume that routing is pre-determined
and, hence, the destination output line of each packet is known. After
service, each packet is moved to its destination output queue. Each
output queue operates using a single queue scheduling policy (eg.
First Come First Serve (FCFS), Weighted Fair Queuing (WFQ) etc.).
The served packets are queued and served according to the single queue
scheduling policy. Fig.~\ref{fig:model} illustrates this model.

Under this setup, the problem of scheduling is to choose a \emph{feasible}
schedule in each time slot to serve the queues and move the packets
to their respective output queues. Since the scheduling policy for
each of the output queues can be chosen independently, the problem
reduces to that of the constrained collection of queues.

\subsubsection{Notation}

First, some general notation. $\mathbb{R}$ denotes the set of real
numbers and $\mathbb{R}_{+}$ the set of non-negative real numbers
i.e., $\mathbb{R}_{+}=\left\{ x\in\mathbb{R}\colon x\geq0\right\} $.
$\mathbb{N}$ denotes the set of natural numbers $\left\{ 1,2,\ldots\right\} $
and $\mathbb{Z}_{+}$ the set of non-negative integers $\left\{ 0,1,2,\ldots\right\} $.
Let $\mathbf{0}$ and $\mathbf{1}$ denote the vectors of $0$s and
$1$s respectively. All the vectors in this paper are length $N$
vectors. Let $\mathbb{}$$\bbone_{\left\{ \cdot\right\} }$ denote
the indicator function, $\bbone_{\text{{true}}}=1$ and $\bbone_{\text{{false}}}=0$.
$x^{+}$ denotes $\max\left\{ x,0\right\} $ and we use the $\ell_{1}$
norm $\left|x\right|=\sum_{n}x_{n}$. We also use the standard inner
product $\dotprod ab=\sum a_{i}b_{i}$.

\begin{figure}
\scalebox{0.9}{\input{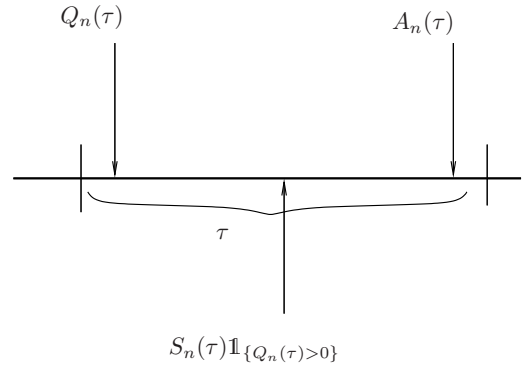}}

\caption{The order in which service and arrivals happen during time slot $\tau$
\label{fig:timeline}}

\end{figure}

Recall that we are assuming Bernoulli i.i.d arrivals. Let $A_{n}(\tau)\in\left\{ 0,1\right\} $
denotes the number of arrivals to queue $n,$ $n=1,2,\ldots,N$, during
time slot $\tau$. The arrival rate to queue $n$ is denoted by $\lambda_{n}$
i.e., $\Pr\left(A_{n}(\tau)=1\right)=\lambda_{n}$ $\forall\tau.$
$\lambda=(\lambda_{n})$ denotes the arrival rate vector. $Q_{n}(\tau)$
denotes the length of queue $n$ at the beginning of time slot $\tau$.
$Q(\tau)=\left(Q_{n}(\tau)\right)$ denotes the queue length vector.
$S(\tau)\in\crS$ denotes the feasible schedule chosen in time slot
$\tau$ to serve the queues. Without loss of generality, we assume
that a feasible schedule is chosen and service happens at the middle
of the time slot, and exogenous arrivals occur at the end of the time
slot. This is shown in Fig.~\ref{fig:timeline}. With $Q_{n}(\tau)$
denoting the queue length at the beginning of time slot $\tau$, we
have: \begin{equation}
Q_{n}(\tau+1)=\left(Q_{n}(\tau)-S_{n}(\tau)\right)^{+}+A_{n}(\tau)\end{equation}

$ $

Finally, let $D_{n}(\tau)$ denote the cumulative departure process
of queue $n$ i.e.,\begin{equation}
D_{n}(\tau)=\sum_{t\leq\tau}S_{n}(t)\bbone_{\left\{ Q_{n}(t)>0\right\} }.\end{equation}

\subsubsection{Definitions}

We now introduce some definitions. We call a system rate stable, or
simply stable in this paper, if the following holds with probability
$1$: for $1\leq n\leq N$,\begin{equation}
\lim_{\tau\to\infty}\frac{{D_{n}(\tau)}}{\tau}=\lambda_{n}\end{equation}

An arrival rate vector $\lambda=\left(\lambda_{n}\right)$ is called
admissible if $\exists$ a scheduling policy under which the queuing
network loaded with $\lambda$ has a queue size process $Q_{n}(\tau)$
such that\begin{equation}
\limsup_{\tau}\E{\abs{Q(\tau)}}<\infty\end{equation}

Let $\Lambda$ denote the set $\left\{ \lambda\in\mathbb{R}_{+}^{N}\colon\lambda\text{\text{ is admissible}}\right\} $.
$\Lambda$ is called the throughput region or capacity region of the
network. Tassiulas and Ephremides (1992) \cite{TE92} proved that:\begin{equation}
\relint\co{\crS}\subseteq\overline{\co{\crS}}\end{equation}

where $\co{\crS}$ denotes the convex hull of $\crS$ i.e., $\left\{ \mu\in\mathbb{R}_{+}^{N}\colon\mu=\sum_{i}\alpha_{i}\pi^{i},\alpha_{i}\geq0,\pi^{i}\in\crS,\sum_{i}\alpha_{i}\leq1\right\} .$
$\relint\co{\crS}$ denotes the relative interior of $\co{\crS}$
and $\overline{\co{\crS}}$ denotes the closure of $\co{\crS}$. Denote
$\relint\co{\crS}$ by $\Lambda'$. It was also shown by Tassiulas
and Ephremides (1992) \cite{TE92} that: 

\begin{multline*}
\Lambda'=\Bigl\{\mu\in\mathbb{R}_{+}^{N}\colon\mu=\sum_{i}\alpha_{i}\pi^{i};\\
\alpha_{i}\geq0,\pi^{i}\in\crS,\sum_{i}\alpha_{i}<1\Bigr\}.\end{multline*}
We call a scheduling algorithm throughput optimal if $\forall\;\lambda\in\Lambda'$,
the system is rate-stable.

\subsubsection{Constraint free network (CFN)}

We now introduce the notion of a Constraint Free Network (CFN). A
constraint free network is defined as a queuing network in which all
the queues can be served simultaneously. Therefore, for a CFN, $\crS=\left\{ 0,1\right\} ^{N}$.
Thus, scheduling just entails moving arriving packets immediately
to their respective destination output queues.

As discussed in Related Work (Section~\ref{sub:Related-work}), fairness
is well understood for single queues. Therefore, using a single queue
fair scheduling scheme for each of the output queues yields a fair
scheduling algorithm for the CFN. We assume throughout that a CFN
is operating using a fair scheduling algorithm. Along with a CFN we
define a shadow CFN as follows: Given a constrained queuing network
$\mathcal{N}$, a CFN $\mathcal{N}'$ with the same number of queues
and fed with copies of exogenous arrivals to $\mathcal{N}$, is called
the shadow CFN of $\mathcal{N}$.

We conclude this section with a brief motivation for the definition
of CFN. This also serves as a preview to our approach to the problem.
As mentioned earlier, the difficulty in designing a fair scheduling
algorithm for networks arises because of the presence of constraints.
In the absence of such constraints, the notion of fairness is equivalent
to that of a single queue. Thus, we define an ideal network that is
constraint-free, whose performance we want to emulate. This is in
some sense the best we can do in providing fairness and, thus, serves
as a benchmark for defining notions of fairness.

\subsection{Scheduling algorithms}

We consider the problem of designing scheduling algorithms for constrained
queuing networks. A scheduling scheme or algorithm is a procedure
whereby an appropriate feasible schedule is chosen in each time slot.
In this paper, we will be interested in a class of scheduling algorithms
termed the Maximum Weight (MW) Scheduling Algorithms. In general,
a maximum weight algorithm works as follows: In each time slot $\tau$,
each queue $n$ is assigned a weight $\omega_{n}(\tau)$. This weight
is usually \textendash{} but not necessarily \textendash{} a function
of the queue size $Q_{n}(\tau)$. Then, the algorithm chooses the
schedule with the maximum weight i.e.,\begin{equation}
S(\tau)=\arg\max_{\pi\in\crS}\dotprod{\omega(\tau)}{\pi}\end{equation}

A feasible schedule $\pi\in\crS$ is said to be maximal if $\forall\sigma\in\crS$,
$\pi\nleq\sigma$ component wise. The set of all maximal feasible
schedules will be denoted by $\crS_{\max}$. It is reasonable to assume
that we want to serve as many queues as possible in each time slot.
Therefore, when the algorithm chooses a feasible schedule $\pi$,
we serve the queues according to a maximal schedule $\mu\in\crS_{\max}$
such that $\mu\geq\pi$.

\noindent \textbf{Remark.} An important special instance of the MW
scheduling algorithm is the one with queue sizes as the weights i.e.,
$\omega_{n}(\tau)=Q_{n}(\tau)$. In their seminal work, Tassiulas
and Ephremides (1992) \cite{TE92} (and independently McKeown et.
al. (1996) \cite{MAW96}) showed that the MW algorithm with queue
sizes as weights is rate stable.

Since these results, there has been a significant work on designing
high-performance, implementable packet scheduling algorithms that
are derivatives of maximum weight scheduling, where weight is some
function of queue-sizes. All of these algorithms are designed to optimize
network utilization as well as minimize delay (for example, see recent
work by Shah and Wischik (2006) \cite{SW06}). However, these algorithms
ignore the requirement of fairness. Specifically, it has been observed
that the maximum weight based algorithm can lead to unwanted starvation
or very unfair rate allocation when switch is overloaded (for example,
see work by Kumar, Pan and Shah (2004) \cite{KPS04}). We provide
a simple example of a switch (detailed description given in the next
subsection) to illustrate this: Consider a 2 \texttimes{} 2 switch
with arrival rate matrix $\lambda_{11}=\lambda_{21}=0$, $\lambda_{12}=0.6,$
$\lambda_{22}=0.5$. Here $\lambda_{ij}$ corresponds to the arrival
rate of traffic at input port $i$ for output port $j$. Under this
loading, output port 2 is overloaded. If OQ switch has Round Robin
(or Fair) policy at output 2 so that traffic from both inputs is served
equally, then input 1 will get rate 0.5 and input 2 will get rate
0.5 from output 2. However, the maximum weight matching policy, with
weight being queue-size (or for that matter any increasing continuous
function of queue-size), the algorithm will try to equalize lengths
of queues at both inputs. Therefore, input 1 will get service rate
0.55 while input 2 will get service rate 0.45 from output 2.

\subsection{Input queued switch}

We now discuss an input queued switch as a special instance of the
abstract model that we have described. As mentioned before, a switch
is present at the core of an Internet router. A router moves packets
from input ports to output ports. Based on the final destination of
the arriving packet, a router determines the appropriate output port
and then transfers the packet accordingly. The transfer of packets
to the corresponding output ports is called switching.

\begin{figure}
\scalebox{0.75}{\input{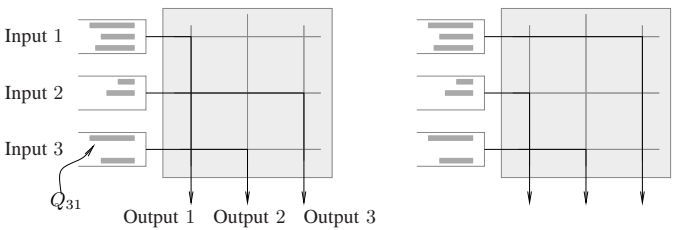}}

\caption{A 3 port input queued switch showing two different possible matchings.
\label{fig:IQswitch}}

\end{figure}

There are various switching architectures, but we discuss the one
that is commercially the most popular. Consider an input queued switch
containing $M$ input ports and $M$ output ports. The queues at the
output ports correspond to the output queues mentioned above, and
hence we retain the notation $M$ for the number of output of queues;
since we are considering only switches with equal number of input
and output ports, the number of input ports is also $M$ . Packets
arriving for input port $i$ and destined for output port $j$ are
stored at input port $i$ in $Q_{ij}$. Note that for a switch, it
is convenient to denote the queues as $Q_{ij}$ instead of as $Q_{n}$,
as we do in the generic model. Further, note that the total number
of queues $N=M^{2}$. The switch transfers packets from input ports
to output ports using the switching fabric. The crossbar switching
fabric implemented in an input queued switch imposes the following
constraints on packet transfer from input to output ports: in each
time slot, each input port can transmit at most one packet and each
output port can receive at most one packet. Therefore, feasible schedules
are matchings from input to output ports. This is illustrated in Fig.~\ref{fig:IQswitch}.
The left and right hand figures illustrate two different possible
matchings.

The scheduling algorithm in the input queued switch chooses an appropriate
matching in each time slot. To link back to the abstract model that
we described, note that an M port switch has $N=M^{2}$ constrained
queues; for the queues we use the notation $\cdot_{ij}$ and not $\cdot_{n}$
for all terms to clearly reference the input and output ports. The
set of all feasible schedules $\crS$ corresponds to the set of all
matchings in an $M\times M$ bipartite graph:

\begin{multline*}
\crS=\Bigl\{\pi=(\pi)_{ij}\in\left\{ 0,1\right\} ^{M\times M}\colon\sum_{k=1}^{M}\pi_{ik}\leq1;\\
\sum_{k=1}^{M}\pi_{kj}\leq1,1\leq i,j\leq M\Bigr\}.\end{multline*}

Packets leave the switch from their respective output ports and hence
the output ports correspond to the output queues. Since at most one
packet arrives at each output port in each time slot, the packet immediately
departs from the output queue. Thus, scheduling reduces to choosing
an appropriate matching in each time slot. We point an interested
reader to \cite{CGMP99} for a more detailed exposition on switch
architectures.

\subsection{Wireless networks}

We now consider wireless networks as a special instance of the abstract
model. Consider a collection of devices (eg. sensor nodes, WiFi nodes,
etc.) that are using the wireless medium to transmit messages. The
devices share the same frequency to transmit and hence interfere when
they transmit simultaneously. Because of power constraints, only devices
that are close to each other geographically can communicate. This
is often modeled by a graph with a node for each device and an edge
between two nodes if they can communicate.

Power constraints also limit interference to nodes that are close
to each other; in other words, only nodes that are connected by an
edge interfere. Therefore, the graph also models the interference
constraints on scheduling. In other words, transmission to a node
is successful only if none of its neighbors in the graph is transmitting
at the same time. This model of interference is termed the \emph{independent
set} model of interference.

\begin{figure}
\scalebox{0.9}{\input{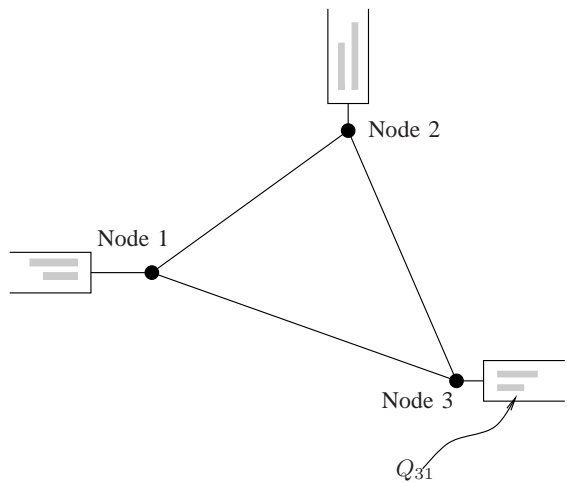}}

\caption{A 3 node wireless network \label{fig:wireless}}

\end{figure}

We assume that the network is modeled as a graph $\mathcal{G}=(\mathcal{V},\mathcal{E})$
with $\mathcal{V}$ denoting the node set $\left\{ 1,2,\dots,N\right\} $
and $\mathcal{E}$ denoting the directed edge set $\left\{ (i,j)\colon i\text{ communicates with }j\right\} $.
Each node $i$ maintains a queue $Q_{ij}$ for each of its neighbors
$j$. We assume a single hop network in which packets arrive at nodes,
get transmitted to one of the neighbors and then leave the network
through output queues. Fig.~\ref{fig:wireless} illustrates a wireless
network with three nodes operating under interference constraints. 

In this setup, the scheduling problem is to decide which directed
links will be active simultaneously. Constraints limit feasible schedules
to those in which none of the neighbors of a receiver is transmitting;
in other words, if link $(i,j)$ is active then none of the links
in the set $\left\{ (l,k)\in\mathcal{E}\colon(l,j)\in\mathcal{E}\text{ OR }(j,l)\in\mathcal{E}\right\} $
should be active. For each network represented by graph $\mathcal{G}=(\mathcal{V},\mathcal{E})$,
we can construct a conflict graph $\mathcal{G}'=(\mathcal{V}',\mathcal{E}')$
with a node for each of the directed links and an edge between two
links if they cannot be active simultaneously. The feasible schedules
then reduce to independent sets in the conflict graph. Formally, \begin{equation}
\crS=\left\{ \pi\in\left\{ 0,1\right\} ^{\abs{\mathcal{V}'}}\colon\pi_{i}+\pi_{j}\leq1,\text{ for all }(i,j)\in\mathcal{E}'\right\} \end{equation}

It should be noted that using the conflict graph, more general constraints
in the network can be modeled as independent set constraints. Thus,
the model we are considering encapsulates the essence of a large class
of scheduling problems. 

\section{Our Approach \label{sec:Our-Approach}}

Network resources are shared by different users and our goal is to
design a scheduling algorithm that allocates resources in a fair manner.
Before we can design such an algorithm, there is a need to give a
precise definition of the users and the resources of the network.
Traditionally, different traffic flows were considered the users and
the bandwidth allocated to them the resource of the network. Link
capacity constraints limited the total amount of resources available,
and each flow was allocated its \textquotedblleft{}fair\textquotedblright{}
share of bandwidth. This is the basis of the \emph{utility maximization}
framework in which the utility of each flow was a function of the
allocated bandwidth \textendash{} the more the bandwidth allocated,
the greater the utility. Different utility functions yield different
fairness criteria. An inherent limitation of this approach is that
it considers entire flows as users, disregarding the fact that flows
are not continuous but are composed of packets. Moreover, bandwidth
is a resource that is allocated on an average over a long period of
time assuming that flow statistics remain constant over such long
periods.

We overcome this limitation by treating the Head-of-Line (HoL) packet
of each flow as the user of the network resources (We assume that
each queue is associated with a flow and hence we use these terms
interchangeably). This takes into account the packetized nature of
a flow and is a consequence of the realization that in each time slot
the decision is whether to serve the HoL packet or not, unlike the
case of a continuous flow that can be given fractional service. Therefore,
utilities should correspond to HoL packets and not entire flows. With
HoL packets as the users, the network resource becomes the service
they receive in each time slot. The network resource is limited by
the constraint set $\crS$ and the algorithm should choose a feasible
schedule $\pi\in\crS$ in a manner that is \textquotedblleft{}fair\textquotedblright{}
to all the HoL packets. Inspired by the utility maximization framework,
we could define utility functions for the HoL packets and choose a
feasible schedule that maximizes the overall utility. But, there is
no natural choice of the utility function and hence we take a different
approach.

We begin with the realization that packets do not have natural utility
functions, but they do have a natural preference order of the feasible
schedules. For each packet, there are two classes of schedules \textendash{}
one class containing all schedules that serve it and the other containing
all schedules that do not. The packet is indifferent to all the schedules
in the same class and the preference relation between schedules in
different classes depends on how \textquotedblleft{}urgently\textquotedblright{}
the packet wants to get served. Fair scheduling now reduces to combining
individual preferences in a fair manner to come up with a \textquotedblleft{}socially\textquotedblright{}
preferred schedule. This is equivalent to a ranked election problem:
HoL packets (queues) are voters, schedules are candidates and each
packet has a preference list of the schedules (refer to Section~\ref{sec:Ranked-Election}
for more details on the ranked election problem). The problem of ranked
election is very well studied in the Economics literature (also called
the theory of social choice). In their seminal work in 1952, Goodman
and Markowitz \cite{GM52} prove that under certain socially desirable
postulates (detailed in Section~\ref{sec:Ranked-Election}), when
the voters have cardinal (quantitative) preferences over candidates,
a simple function of those quantitative preferences yields a uniquely
preferred outcome.

In order to use the Goodman and Markowitz result, we require not just
relative preferences, but quantitative preferences over the feasible
schedules. By quantitative preferences we mean that each packet assigns
numerical weights to schedules; the higher the weight, the more preferred
the schedule. In principle, the packet can assign any weights consistent
with its relative order of preferences to obtain quantitative preferences
over the schedules. But it is important to realize that the choice
of quantitative preference is crucial for obtaining the practically
desired benefits of the fair scheduling. In our setup, each packet
has two classes of schedules: one that it prefers to the other while
being indifferent to schedules within the same class. Therefore, the
packet assigns the same weight to all the schedules within the same
class. Since only relative preferences matter, we assume that each
packets assigns the same weight $0$ to all schedules in the class
it does not prefer. Assigning quantitative preferences now reduces
to choosing a weight for each packet to assign to the class of schedules
it prefers.

One feasible option would be to use queue sizes as weights. The problem
with this choice is that it is oblivious to flow identities and is
susceptible to manipulation (a flow can gain advantage by overloading
system with packets resulting in large queue sizes). Another option
would be to use the age (waiting time in the system) of the packet.
This choice is still oblivious to flow and packet identities and it
is difficult to provide QoS by giving priority to one flow (packet)
over other . We overcome these problems by using the idea of emulation
of the shadow CFN. As mentioned in Related Work (Section~\ref{sub:Related-work}),
one way of designing a fair scheduling algorithm would be to perfectly
emulate a CFN using a fair queuing policy at each of the output queues.
But this results in a loss of throughput of the system. Therefore,
our approach would be to emulate the shadow CFN as closely as possible.
In this spirit, we use a function of the departure time of the packet
from the shadow CFN as the weight; the earlier the departure time,
the higher the weight. The details of the exact function used are
covered in Sections~\ref{sec:Analogy-between-fair} and \ref{sec:Most-Urgent-Cell}.

We now tie this back to the utility maximization framework. Using
the Goodman and Markowitz algorithm with the above choice of weights
yields a MW style algorithm with the weight of each queue equal to
the weight assigned to the packet. This is identical to the result
we obtain by using the assigned weights as utilities of packets and
choosing a schedule that maximizes overall utility. Therefore, our
algorithm yields utility functions for packets that can be used in
the utility maximization framework. This rather surprising result
connects our approach back to utility maximization very nicely.

We then establish that such an algorithm is throughput optimal under
the standard stochastic model of a network. To prove throughput optimality
(rate stability to be precise), we use an appropriate quadratic Lyapunov
function. However, we cannot use the standard stability proof technique
based on Foster's criterion because the Lyapunov function is not a
function of queue-sizes, but is function of \emph{preferences }derived
from the shadow CFN. This makes the analysis rather non-trivial.

To explain the consequences of our algorithm on fair emulation, we
present simulations for algorithms based on FIFO OQ switch. Intuitively,
our fair algorithm should be able to reduce the queue-size (or delay)
as well as get rid of starvation caused by well-known throughput optimal
algorithms. Our simulation results clearly confirm this intuition.

\section{Ranked Election\label{sec:Ranked-Election}}

In this section we take a digression into Economics literature to
describe the ranked election problem. 
\begin{defn}
[Ranked election] There are $M$ voters that vote for $C$ candidates.
Vote of each voter consists of a ranking (or permutation) of all $C$
candidates. These votes can additionally carry quantitative values
associated with their preferences. Let $a_{mc}$ denote the value
voter $m$ gives to candidate $c$, for $1\leq m\leq M$, $1\leq c\leq C$.
The goal of the election is to relative order all the $C$ candidates
as well as produce the ultimate winner in a manner that is consistent
with the votes.
\end{defn}
The key for a good election lies in defining consistency of the outcome
of election with votes. The following are canonical postulates that
are used in the literature on ranked election:

\begin{itemize}
\item[P1.] Between any two candidates $c$ and $c'$, suppose that none of the $M$ voters prefers $c'$ to $c$ and at least one voter prefers $c$ to $c'$. Then $c'$ should not be ranked higher than $c$ in the output of the election. This property corresponds to the economic notion of weak Pareto optimality.
\item[P2.] Suppose the voters are renumbered (or renamed) while keeping their votes the same. Then the outcome of election should remain the same. In other words, the election outcome is blind to the identity of the voters, that is election outcome is symmetric.
\item[P3.] Now, consider the setup when the votes are cardinal (i.e., quantitative). Suppose candidate $c$ is preferred to $c'$ by the election. Then, by adding the same fixed constant to all $a_{mc}$ and $a_{mc'}$ for $1 \leq m \leq M$, the relative order of candidates $c$ and $c'$ should not change. This makes sense because what matters is the difference in preference levels for the two candidates, not the actual values.
\end{itemize}

In the absence of cardinal (or quantitative) preferences, the question
of ranked election with postulates P1, P2 (and some additional postulates)
was first studied by Arrow (1951)~\cite{A51}. In his celebrated
work, he established the (then) very surprising impossibility of the
existence of any election scheme that satisfies P1, P2 (and additional
postulates) simultaneously. We note that this result has been an important
corner stone in the field of theory of social choice.

Subsequent to Arrow's impossibility result, many economists started
looking for positive results. Among many other celebrated results,
the result that is relevant to this paper is that of Goodman and Markowitz
(1952)~\cite{GM52}. They showed that if voters have cardinal preferences,
as in our setup, then there is a unique ordering of candidates that
satisfies P1-P2-P3 simultaneously. To describe their result, consider
the following: let the net score of a candidate$c$ be $s_{c}=\sum_{m=1}^{M}a_{mc}$.
Goodman and Markowitz obtained the following remarkable result.
\begin{thm}
\label{thm:GM}Suppose the scores of all candidates are distinct.
Rank candidates as follows: candidate $c$ has higher ranking than
$c\lyxmathsym{\textasciiacute}$ if and only if $s_{c}>s_{c}'$. This
ranking satisfies postulates P1-P2-P3. Further, this is the only such
ranking.
\end{thm}
For a proof of this result, we refer the reader to~\cite{GM52}.

\section{Analogy between fair scheduling and ranked election\label{sec:Analogy-between-fair}}

In this section, we motivate our fair scheduling algorithm by establishing
an equivalence between fair scheduling in a constrained queuing network
$\mathcal{N}$ and the problem of ranked election. In our context,
the packets (queues) are the voters and the feasible schedules $\pi\in\crS$
are the candidates. In order to use the Goodman and Markowitz setup,
we need to derive preferences for packets over schedules. For each
packet, there are two classes of schedules \textendash{} one class
containing all schedules that serve it and the other containing all
schedules that do not. The packet is indifferent to all the schedules
in the same class. Since only the relative weights matter, we assume
that a packet assigns a weight of $0$ to all schedules that do not
serve it.

We derive preferences for packets over schedules that serve them from
the corresponding shadow CFN $\mathcal{N}\lyxmathsym{\textasciiacute}$
that is operating with a single queue fair scheduling policy at each
of its output queues. As defined before, a copy of every packet arriving
to the network $\mathcal{N}$ is fed to the shadow CFN $\mathcal{N}\lyxmathsym{\textasciiacute}$.
That is, (a copy of) a packet arriving at queue n for output queue
$m$ of $\mathcal{N}$ immediately joins the output queue $m$ in
$\mathcal{N\lyxmathsym{\textasciiacute}}$. The departures from the
output queues of $\mathcal{N\lyxmathsym{\textasciiacute}}$ happen
according to an appropriate fair scheduling policy, say $\mathcal{P}$,
such as strict priority scheme, last-in-first-out or simply first-in-first-out.
Specifically, our objective in assigning preferences is to have the
departures of packets from $\mathcal{N}$ be as close as possible
to the departures from the corresponding shadow CFN $\mathcal{N\lyxmathsym{\textasciiacute}}$.
Ideally, we want $\mathcal{N}$ to exactly emulate $\mathcal{N\lyxmathsym{\textasciiacute}}$
i.e., we want the departure times of packets from both the networks
to be exactly the same. However, we settle with approximate emulation
because, as shown by Chuang et. al. \cite{CGMP99}, exact emulation
is not possible at speedup 1. Since the preferences of packets are
chosen from $\mathcal{N\lyxmathsym{\textasciiacute}}$ and these preferences
are combined in a fair manner, the fair scheduling polices at the
output queues of $\mathcal{N\lyxmathsym{\textasciiacute}}$ can now
be chosen according to the desired requirements. 

Based on the above discussion, our approach is to use a value that
is a function of the departure time of the packet from$\mathcal{N\lyxmathsym{\textasciiacute}}$
\textendash{} the earlier the departure time, the higher the value
assigned. More specifically, let $p_{n}^{\tau}$ denote the HoL packet
in queue $n$ of network $\mathcal{N}$ at time $\tau$. Let $d_{n}(\tau)$
denote the departure time of $p_{n}^{\tau}$ from $\mathcal{N\lyxmathsym{\textasciiacute}}$.
For the following discussion we assume that the queue is non-empty
and hence $d_{n}(\tau)$ is well defined. We defer the discussion
of empty queues to the next section. Now, each queue $n$ assigns
a value of $\tau-d_{n}(\tau)$ to all the schedules that serve it.
(The choice of $\tau-d_{n}(\tau)$ seems arbitrary, when we could
have taken any decreasing function of $d_{n}(\tau)$. Indeed we can,
though it should have some \textquotedblleft{}nice\textquotedblright{}
properties to maximize throughput. Details are in Section~\ref{sec:Throughput-of-MUCF}).
This completes the equivalence.

Taking a closer look at the weight $\tau-d_{n}(\tau)$, note the following.
Suppose at time $\tau$ the packet $p_{n}^{\tau}$ is already late
i.e, $d_{n}(\tau)<\tau$. In this case, the weight $\tau-d_{n}(\tau)>0$,
which means that $p_{n}^{\tau}$ prefers all schedules that serve
it to all the schedules that do not by weight $\tau-d_{n}(\tau)$.
Thus, the more delayed the packet is, the higher the weight it assigns.
On the other hand, when the packet is not late i.e., $d_{n}^{\tau}>\tau$,
$p_{n}^{\tau}$ prefers schedules that do not serve it in order to
give a chance to packets that are late to get served.

Now, with the above assignment of values to each schedule by each
queue, the value of a schedule $\pi\in\crS$ is given as: \begin{align*}
\v{\pi} & =\sum_{n=1}^{N}\pi_{n}(\tau-d_{n}(\tau))\\
 & =\dotprod{\tau-d(\tau)}{\pi}\end{align*}

The postulates P1-P2-P3 translate into the following postulates for
network scheduling.

\begin{itemize}
\item[P1$'$.] Between any two schedules $n_1$ and $n_2$, suppose that none of the $N$ HoL packets prefer $n_2$ to $n_1$ and at least one HoL packet prefers $n_1$ to $n_2$. Then, we should not choose $n_2$.
\item[P2$'$.] For given HoL packets, let $\pi$ be the outcome of the election as per the above preferences for schedules. Then, by renumbering queues while retaining the same HoL preferences, the outcome of election should be only renumbered $\pi$. In other words, the election does not give unfair priority to any port and thus is symmetric in its inputs.
\item[P3$'$.] Suppose schedule $\pi^1$ is preferred to $\pi^2$ by the election. By adding the same fixed constant to $\tau-d_n(\tau)$ for all $n$, the outcome of the election should remain unchanged.
\end{itemize}

The election algorithm of Goodman and Markowitz suggests that the
following schedule $S(\tau)$ should be chosen:

\[
S(\tau)\in\arg\max_{\pi\in\crS_{\max}}\dotprod{\tau-d(\tau)}{\pi}\]

\section{Most Urgent Cell First (MUCF($f$)) Algorithm\label{sec:Most-Urgent-Cell}}

Based on the discussion in the previous section, we propose a fair
scheduling algorithm called the \emph{most urgent cell first} algorithm.
According to this algorithm, packets are scheduled according to the
maximum weight schedule (MWS) with urgencies of queues as weights
with the urgency $U_{n}(\tau)$ of queue n defined as $\tau-d_{n}(\tau)$
if it is non-empty i.e., the more \textquotedblleft{}late\textquotedblright{}
the packet is, the higher is its urgency. If queue $n$ is empty we
define its urgency as $-\max\left\{ 0,-\min_{n\colon Q_{n}(\tau)>0}U_{n}(\tau)\right\} $.
Note that according to this definition, the weight assigned to a schedule
by an empty queue is always less than or equal to the weight assigned
by any non-empty queue in the system. Therefore, as desired, the fair
schedule chosen by the algorithm tries to minimize the number of empty
queues it serves. Of course, as mentioned in the previous section,
we could have used any increasing function of the urgency. In particular,
suppose $f\colon\mathbb{R}\to\mathbb{R}$ denotes a non-decreasing
bi-Lipschitz continuous function with Lipschitz constant $\rho>1$
i.e., for any $x,y\in\mathbb{R}$, $1/\rho\abs{x-y}\leq\abs{f(x)-f(y)}\leq\rho\abs{x-y}$.
Without loss of generality, we assume that $f(0)=0$; thus, $f(x)\geq0$
for $x\geq0$. The MUCF($f$ ) algorithm now chooses the MWS with
weight of queue $n$ equal to $U_{n}^{f}(\tau)$, defined as $f(U_{n}(\tau))$.
Formally, the three components of the algorithm are as follows:

\begin{minipage}[t]{1\columnwidth}%
\begin{shaded}%
\begin{enumerate}
\item The arriving packets are queued according to the \textbf{FIFO} queuing
policy in each of the N constrained queues. 
\item For a packet $p$ in $\mathbb{\mathcal{N}}$ , let $d(p)$ denote
its departure time from $\mathcal{N\lyxmathsym{\textasciiacute}}$.
Then, the arriving packets in each of the $M$ output queues are queued
in the order of their respective departure times from $\mathcal{N\lyxmathsym{\textasciiacute}}$.
More formally, in every output queue $m$, a packet $p$ will be ahead
of every packet $p\lyxmathsym{\textasciiacute}$ that satisfies $d(p\lyxmathsym{\textasciiacute})>d(p)$.
\item In each time slot $\lyxmathsym{\textgreek{t}}$ , the algorithm chooses
a feasible schedule $S(\tau)$ from $\crS$ using a MW criterion as
follows: \begin{equation}
S(\tau)\in\arg\max_{\pi\in\crS_{\max}}\dotprod{U^{f}(\tau)}{\pi}\end{equation}

\end{enumerate}
\end{shaded}%
\end{minipage}

\section{Throughput of MUCF($f$) Algortihm\label{sec:Throughput-of-MUCF}}

The previous section described how we arrived at MUCF algorithm as
a fair algorithm based on preferences obtained from a shadow CFN.
As established in the previous section, Theorem~\ref{thm:GM} implies
that MUCF is the only algorithm that satisfies the desirable postulates
P1$'$-P2$'$-P3$'$. In this section, we state and prove the throughput
optimality property of the MUCF algorithm. The proof of the algorithm
is non-traditional and requires new techniques that may be of interest
for analysis of such non-queue based weighted algorithms.
\begin{thm}
\label{thm:rate-stable}Consider a constrained queuing system with
an arbitrary set of constraints $\crS$ . Suppose the system is loaded
with an i.i.d. Bernoulli arrival process, and is operating under the
MUCF($f$) algorithm with$f(\cdot)$ a bi-Lipschitz function. Then,
if the rate vector is strictly admissible, the queuing network is
rate stable.
\end{thm}
Before we prove Theorem~\ref{thm:rate-stable} we need the following
notation and lemmas.

\noindent \textbf{Notation.} First, some useful notation. Consider
the HoL packet $p_{n}^{\tau}$ of queue $n$ in network $\mathcal{N}$
at the beginning of the time slot $\lyxmathsym{\textgreek{t}}$ .
As before, let $a_{n}(\tau)$ be its time of arrival and $d_{n}(\tau)$
be the time of its departure from $\mathcal{N\lyxmathsym{\textasciiacute}}$,
$U_{n}(\tau)$ be its urgency as defined above, and $W_{n}(\tau)$
be its waiting time (i.e., $\tau-a_{n}(\tau)$ if the queue is non-empty
and $0$ if it is empty). Let $W_{n}^{f}(\tau)$ denote $f(W_{n}(\tau))$
and $F(y)$ denote $\int_{o}^{y}f(x)dx$. Also, define $\Delta_{n}(\tau)$
as $W_{n}(\tau)-U_{n}(\tau)$. We note that if queue $n$ is empty,
then $W_{n}(\tau)=0$ and $U_{n}(\tau)$ is as defined above. Hence,
$\Delta_{n}(\tau)$ is always non-negative. Let $B_{m}^{k}$, $m=1,2,\ldots,M$,
denote the length of the $k^{\mbox{th}}$ busy cycle at output queue
$m$ in $\mathcal{N\lyxmathsym{\textasciiacute}}$. Finally, for any
function $g\colon\mathbb{R}\to\mathbb{R}$ and a vector $v\in\mathbb{R}^{N}$,
$g(v)\in\mathbb{R}^{N}$ denotes $(g(v_{n}))$. Before we state the
lemmas and prove Theorem~\ref{thm:rate-stable}, note the following
property of $F(\cdot)$:\begin{equation}
F(0)=0,\quad F(y)\geq0,\;\mbox{for all }y\in\mathbb{R}.\label{eq:12}\end{equation}

The equality $F(0)=0$ follows directly from the definition of $F(\cdot)$.
Coming to $F(y)\geq0$, note that since $f(\cdot)$ is non-decreasing
and $f(0)=0$, it follows that $f(x)\leq0$ for $x\leq0$ and $f(x)\geq0$
for $x>0$. Hence, for $y>0$, $F(y)=\int_{0}^{y}f(x)dx\geq0$ since
$f(x)\geq0$ for $x>0$. Similarly, for $y<0$, $F(y)=\int_{0}^{y}f(x)dx=\int_{y}^{0}(-f(x))dx\geq0$
since $-f(x)\geq0$ for $x<0$. 
\begin{lem}
\label{lem:lyapunov}Let $L(\tau)\colon=\dotprod{F(W(\tau))}{\lambda}=\sum_{n}F(W_{n}(\tau))\lambda_{n}$.
Then, under the MUCF($f$) algorithm with $f(\cdot)$ a bi-Lipschitz
function and $\lambda$ being strictly admissible, there exists an
$\varepsilon>0$ such that\[
\E{L(\tau+1)-L(\tau)}\leq-\varepsilon\E{\abs{W(\tau)}}+2\E{\abs{\Delta(\tau)}}+K,\]

for some constant $K$.
\end{lem}

\begin{lem}
\label{lem:W}Under the MUCF($f)$ algorithm, with $f(\cdot)$ a bi-Lipschitz
function and $\lambda$ being strictly admissible, suppose $Z_{\tau}=\frac{1}{\tau\log\tau}\sum_{i=1}^{\tau}\abs{W(i)}$
and $\E{Z_{\tau}}\leq O(1)<\infty$ for all $\tau.$ Then, we must
have\begin{equation}
\Pr\left(\lim_{\tau\to\infty}\frac{1}{\tau}\abs{W(\tau)}=0\right)=1.\label{eq:211}\end{equation}

\end{lem}

\begin{lem}
\label{lem:WCP}Let $\Theta_{m}(\tau)$ denote $\max_{0\leq k\leq\tau}B_{m}^{k}$,
for $m=1,2,\ldots,M$. Then, under a strictly admissible $\lambda$
with the output queues of $\mathcal{N\lyxmathsym{\textasciiacute}}$
operating under a Work Conserving Policy (WCP), the following is true
for all $t$and $1\leq m\leq M$,\[
\E{\Theta_{m}(\tau)}\leq O(\log\tau)\]

\end{lem}
We will first prove the result of Theorem~\ref{thm:rate-stable}
assuming the results of Lemmas~ \ref{lem:lyapunov} and \ref{lem:WCP},
and defer their proof until after the proof of Theorem~\ref{thm:rate-stable}.
\begin{IEEEproof}
[Proof of Theorem~\ref{thm:rate-stable}]We first note that if queue
$n$ is non-empty then \begin{equation}
\Delta_{n}(\tau)\leq\max_{0\leq k\leq\tau}B_{m}^{k}\label{eq:delta-bound}\end{equation}
where $m$ is the destination output queue of packet $p_{n}^{\tau}$.
This is true because when queue $n$ is non-empty, $\Delta_{n}(\tau)$
denotes the waiting time of $p_{n}^{\tau}$ in its destination output
queue in $\mathcal{N\lyxmathsym{\textasciiacute}}$, and hence cannot
be more than the length of the busy cycle it belongs to. Since there
can be at most $\tau$ busy cycles up to time $\tau$ and $p_{n}^{\tau}$
arrived to $\mathcal{N\lyxmathsym{\textasciiacute}}$ before $\tau$,
(\ref{eq:delta-bound}) should be true. Therefore, from lemma~\ref{lem:WCP}
and (\ref{eq:delta-bound}) it follows that:\begin{equation}
\E{\Delta_{n}(\tau)}\leq O(\log\tau)\label{eq:delta-bound2}\end{equation}

If queue $n$ is empty, then by definition it follows that either
$\Delta_{n}(\tau)=0$ or $\Delta_{n}(\tau)=d_{n'}(\tau)-\tau$, for
some queue $n'$ that is non-empty. Since, $\Delta_{l}(\tau)\geq0$
$\forall l,\tau$, we have from (\ref{eq:delta-bound2}) that $\E{\Delta_{n}(\tau)}\leq\E{\Delta_{n'}(\tau)}\leq O(\log\tau)$.
Hence, (\ref{eq:delta-bound2}) is valid even for empty queues and
it thus follows that:\begin{equation}
\E{\abs{\Delta_{n}(\tau)}}\leq O(\log\tau)\label{eq:delta-bound3}\end{equation}

From lemma~\ref{lem:lyapunov} and (\ref{eq:delta-bound3}), we obtain
the following:\begin{equation}
\E{L(\tau+1)-L(\tau)}\leq-\varepsilon\E{\abs{W(\tau)}}+O(\log\tau)+K,\label{eq:16}\end{equation}

Telescopic summation of (\ref{eq:16}) from $1,2,\ldots,\tau$, we
obtain (after cancellations),\begin{multline}
\E{L(\tau+1)}\leq\E{L(0)}-\varepsilon\E{\sum_{i=1}^{\tau}\abs{W(i)}}\\
+O(\tau\log\tau)+\tau K,\label{eq:17}\end{multline}

Now, the network starts empty at time $0$. Therefore, $\E{L(0)}=0$.
Further, $L(\cdot)$ is non-negative function. Therefore, (\ref{eq:17})
gives us 

\begin{equation}
\varepsilon\E{\sum_{i=1}^{\tau}\abs{W(i)}}\leq O(\tau\log\tau)+\tau K.\end{equation}

Dividing both sides by $\varepsilon\tau\log\tau,$ we obtain\begin{equation}
\E{\frac{1}{\tau\log\tau}\sum_{i=1}^{\tau}\abs{W(i)}}\leq O(1).\label{eq:19}\end{equation}

Let $X_{\tau}=\frac{1}{\tau}\sum_{i=1}^{\tau}\abs{W(i)}$ and $Z_{\tau}=\frac{X_{\tau}}{\log\tau}$.
From (\ref{eq:19}), we have $\E{Z_{\tau}}\leq O(1)<\infty$ for all
$\tau$. It now follows from Lemma~\ref{lem:W} that\begin{equation}
\Pr\left(\lim_{\tau}\frac{1}{\tau}\abs{W(\tau)}=0\right)=1\label{eq:20}\end{equation}
Using (\ref{eq:20}), we complete the proof of rate stability of the
algorithm as follows. At time $\tau$, the waiting time of the HoL
packet of queue $n$ is $W_{n}(\tau)$. Because of FIFO policy and
at most one arrival per time slot, we have that the queue-size of
queue $n$ at time $\tau$, $Q_{n}(\tau)\leq W_{n}(\tau)$. From (\ref{eq:20}),
we have that\begin{equation}
\lim_{\tau\to\infty}\frac{Q_{n}(\tau)}{\tau}=0,\mbox{ with probability }1.\label{eq:21}\end{equation}
Now, $Q_{n}(\tau)$ observes the following dynamics: \begin{equation}
Q_{n}(\tau)=Q_{n}(0)+\sum_{t\leq\tau}A_{n}(t)-D_{n}(\tau),\label{eq:22}\end{equation}
where the second term on RHS is the cumulative arrival to queue $n$
till time $\tau$ while the third term is the cumulative departure
from queue $n$ till time $\tau$. By strong law of large numbers
(SLLN) for Bernoulli i.i.d. process we have that: \[
\lim_{\tau\to\infty}\frac{1}{\tau}\sum_{t\leq\tau}A_{n}(t)=\lambda_{n}.\]

Using this along with (\ref{eq:21}) and (\ref{eq:22}), we obtain\[
\lim_{\tau\to\infty}\frac{D_{n}(\tau)}{\tau}=\lambda_{n},\quad\mbox{with probability 1,}\:\forall n.\]
This completes the proof of Theorem~\ref{thm:rate-stable}.
\end{IEEEproof}

\begin{IEEEproof}
[Proof of Lemma~\ref{lem:W}] Suppose (\ref{eq:211}) is not true.
Then, since $\abs{W(\tau)}\geq0$ we have that for some $\delta>0$,\begin{equation}
\Pr\left(\abs{W(\tau)}\geq\delta\tau,\mbox{ i.o.}\right)\geq\delta,\label{eq:23}\end{equation}
where \textquotedblleft{}i.o.\textquotedblright{} means infinitely
often. Now if $\abs{W(\tau)}\geq\delta\tau$, then there exists an
HoL packet that has been waiting in the network for time at least
$\delta\tau/N$. This is true because $\abs{W(\tau)}$ is the sum
of waiting times of at most $N$ HoL packets. Call this packet $p$.
This packet must have arrived at time $\leq\tau-\delta\tau/N=\tau(1-\delta)N^{-1}$.
Since waiting time of a packet increases only by $1$ each time-slot,
the waiting time of packet $p$ must be at least $0.5\delta\tau/N$
in time interval $[\tau_{1},\tau]$, where $\tau_{1}=\tau-0.5\delta\tau N^{-1}=\tau(1-0.5\delta N^{-1}).$
Now, consider any time $\tau'\in[\tau_{1,}\tau]$. The packet waiting
at the HoL of the queue that contains $p$ must have waiting time
higher than that of $p$ due to the FIFO ordering policy. Therefore,
the contribution to $\abs{W(\tau')}$ by HoL packets of the queue
that contains packet $p$ is at least $0.5\delta\tau N^{-1}$. Therefore,
we obtain\begin{equation}
\sum_{\tau'=\tau_{1}}^{\tau}\abs{W(\tau')}\geq(\tau-\tau_{1})\frac{\delta\tau}{2N}=\frac{\delta^{2}\tau^{2}}{4N^{2}}\label{eq:24}\end{equation}
Therefore, by the definition of $X_{\tau}$ and non-negativity of
$\abs{W(\cdot)}$, we have the following logical implication:\begin{equation}
\abs{W(\tau)}\geq\delta\tau\implies X_{\tau}\geq\frac{\delta^{2}\tau}{4N}\label{eq:25}\end{equation}
Thus, if (\ref{eq:23}) holds then by (\ref{eq:25}) we have\begin{equation}
\Pr\left(X_{\tau}\geq\frac{\delta^{2}\tau}{4N^{2}},\mbox{ i.o. }\right)\geq\delta.\label{eq:26}\end{equation}
Now observe the following relation of $X_{t}$: since $\abs{W(\cdot)}\geq0,$\[
X_{t+1}\geq\left(1-\frac{1}{t+1}\right)X_{t}.\]
For any integer $L>0$ and any integer $t'\in[t,t+L]$, we can now
write\begin{align*}
X_{t+L}\geq\prod_{i=t'+1}^{t+L}\left(1-\frac{1}{i}\right)X_{t'} & \geq\prod_{i=t+1}^{t+L}\left(1-\frac{1}{i}\right)X_{t'}\\
 & \geq\left(1-\frac{1}{t}\right)X_{t'}.\end{align*}
Now, for any $\alpha>1$, let $L=\left\lfloor t\alpha\right\rfloor -t$,
which implies that $L\leq(\alpha-1)t$. We can then write for any
integer $t'\in[t,\alpha t]$,\[
X_{\left\lfloor t\alpha\right\rfloor }\geq\left(1-\frac{1}{t}\right)^{L}X_{t'}\geq\left(1-\frac{1}{t}\right)^{(\alpha-1)t}X_{t'}.\]
Since\[
\left(1-\frac{1}{t}\right)^{(\alpha-1)t}\approx\exp(-\alpha+1),\]
taking $\alpha=1.5$, for $t$ large enough, it follows that $\left(1-\frac{1}{t}\right)^{(\alpha-1)t}\approx\exp(-1/2)\geq1/2$.
Thus, for $t$ large enough and any integer $t'\in[t,1.5t]$\begin{equation}
X_{\left\lfloor 1.5t\right\rfloor }\geq\frac{1}{2}X_{t'}\label{eq:27}\end{equation}
Define $Y_{k}=X_{\lfloor1.5^{k}\rfloor}$ for $k\geq0$. Then, the
following are direct implications of (\ref{eq:27}): for any $\theta>0$,\begin{gather}
X_{\tau}\geq\theta\tau\;\mbox{i.o.}\;\implies Y_{k}\geq\theta1.5^{k}/3,\;\mbox{i.o.;}\nonumber \\
Y_{k}\geq\theta1.5^{k},\;\mbox{i.o.}\;\implies X_{\tau}\geq\theta\tau,\;\mbox{i.o..}\label{eq:28}\end{gather}
The first implication is true because for any $\tau$ such that $X_{\tau}\geq\theta\tau,$we
can find a $k$ such that $\tau\in[1.5^{k-1},1.5^{k}].$ It then follows
from (\ref{eq:27}) that $Y_{k}=X_{\lfloor1.5^{k}\rfloor}\geq1/2X_{\tau}\geq\theta\tau/2\geq\theta1.5^{k-1}/2.$
Similarly, for the second implication, whenever $Y_{k}\geq\theta1.5^{k},$
taking $\tau=\left\lfloor 1.5^{k}\right\rfloor $, we can write $X_{\tau}\geq\theta1.5^{k}\geq\theta\tau.$

It follows from (\ref{eq:28}) that $\Pr\left(X_{\tau}\geq\theta\tau\right)\leq\Pr\left(Y_{k}\geq\theta1.5^{k}/3\right).$
Thus, in order to complete the proof of (\ref{eq:20}) by contradicting
(\ref{eq:26}), and thereby (\ref{eq:23}), it is sufficient to show
that for $\theta=\delta^{2}/(4N^{2})$,\[
\Pr\left(Y_{k}\geq\theta1.5^{k}/3,\:\mbox{i.o.}\right)=0.\]
For this, let event $E_{k}=\left\{ Y_{k}\geq\theta1.5^{k}/3\right\} .$
Then, from $\E{Z_{t}}\leq O(1),$ relations $Y_{k}=X_{1.5^{k}},$
$Z_{k}=\frac{X_{k}}{\log k}$ and Markov's inequality we obtain that\[
\Pr\left(E_{k}\right)\leq\frac{3\E{Y_{k}}}{\theta1.5^{k}}=\frac{3\log\left\lfloor 1.5^{k}\right\rfloor \E{Z_{\left\lfloor 1.5^{k}\right\rfloor }}}{}\leq O\left(\frac{k}{1.5^{k}}\right).\]
Therefore, \[
\sum_{k}\Pr\left(E_{k}\right)\leq\sum_{k}O\left(\frac{k}{1.5^{k}}\right)<\infty.\]
Therefore, by Borel-Cantelli's Lemma, we have that \[
\Pr\left(E_{k}\;\mbox{i.o.}\right)=0.\]
This completes the proof of the lemma.
\end{IEEEproof}

\begin{IEEEproof}
[Proof of Lemma~\ref{lem:lyapunov}] Define the following: for all
$n$\[
\tilde{W_{n}}(\tau+1)=W_{n}(\tau)+1-\beta_{n}S_{n}^{*}(\tau)\]
where $S_{n}^{*}(\tau)$ is the schedule of MUCF algorithm and $\beta_{n}$
is the inter-arrival time for the arrival process to queue $n.$ When
the queue is empty, treat $\beta_{n}$ as an independent r.v. without
any meaning, while if queue is not empty then treat it as the inter-arrival
time between the packet being served and the packet behind it. In
either case, due to the FIFO policy $\beta_{n}$ is totally independent
of the scheduling decisions performed by the algorithm till (and including)
time $\tau$ and the information utilized by the algorithm. Therefore,
we will treat it as an independent random variable with Geometric
distribution of parameter $\lambda_{n}$ (since arrival process is
Bernoulli i.i.d.). Consider the following: for any $\tau$,\begin{equation}
F\Bigl(\tilde{W_{n}}(\tau+1)\Bigr)-F\Bigl(W_{n}(\tau)\Bigr)=\int_{W_{n}(\tau)}^{\tilde{W_{n}}(\tau+1)}f(x)dx\label{eq:29}\end{equation}
It is easy to see that,\begin{equation}
\int_{W_{n}(\tau)}^{\tilde{W_{n}}(\tau+1)}f(x)dx=\int_{0}^{1-\beta_{n}S_{n}^{*}(\tau)}f\left(y+W_{n}(\tau)\right)dy\label{eq:30}\end{equation}
Since $f(\cdot)$ is non-decreasing bi-Lipschitz continuous with $f(0)=0,$
we have \begin{align}
f\left(y+W_{n}(\tau)\right) & =f\left(y+W_{n}(\tau)\right)-f\left(W_{n}(\tau)\right)+f\left(W_{n}(\tau)\right)\nonumber \\
 & \leq\abs{f\left(y+W_{n}(\tau)\right)-f\left(W_{n}(\tau)\right)}+W_{n}^{f}(\tau)\nonumber \\
 & \leq\rho\abs y+W_{n}^{f}(\tau)\label{eq:31}\end{align}
Now, it follows from (\ref{eq:29}), (\ref{eq:30}), and (\ref{eq:31})
that\begin{align}
 & F\Bigl(\tilde{W_{n}}(\tau+1)\Bigr)-F\Bigl(W_{n}(\tau)\Bigr)\nonumber \\
\leq\; & \rho\left(1-\beta_{n}S_{n}^{*}(\tau)\right)^{2}+\left(1-\beta_{n}S_{n}^{*}(\tau)\right)W_{n}^{f}(\tau)\label{eq:32}\end{align}
Using (\ref{eq:32}) and the fact that $\beta_{n}$ is a Geometric
r.v. with mean $1/\lambda_{n},$ we have the following:\begin{align}
 & \E{\sum_{n}\lambda_{n}F\Bigl(\tilde{W_{n}}(\tau+1)\Bigr)-\sum_{n}\lambda_{n}F\Bigl(W_{n}(\tau)\Bigr)\biggl|W(\tau)}\nonumber \\
\leq\; & \sum_{n}W_{n}^{f}(\tau)\lambda_{n}-\sum_{n}W_{n}^{f}(\tau)S_{n}^{*}(\tau)+\rho\sum_{n}\lambda_{n}\nonumber \\
 & -2\rho\sum_{n}S_{n}^{*}(\tau)+2\rho\sum_{n}S_{n}^{*}(\tau)\lambda_{n}^{-1}\label{eq:33}\end{align}
Here we have used the fact that $S_{n}^{*}(\tau)\in\left\{ 0,1\right\} $
and hence $\left(S_{n}^{*}(\tau)\right)^{2}=S_{n}^{*}(\tau)$, and
$\E{\beta_{n}^{2}}=2/\lambda_{n}^{2}-1/\lambda_{n}\leq2/\lambda_{n}^{2}$.
Using the fact that $\sum_{n}\lambda_{n}\leq N,$ $\sum_{n}S_{n}^{*}(\tau)\leq N$
and $\lambda_{n}^{-1}<\infty$ for all $n$ such that $\lambda_{n}\neq0$,
we obtain \begin{align}
 & \E{\sum_{n}\lambda_{n}F\Bigl(\tilde{W_{n}}(\tau+1)\Bigr)-\sum_{n}\lambda_{n}F\Bigl(W_{n}(\tau)\Bigr)\biggl|W(\tau)}\nonumber \\
\leq\; & \dotprod{W_{n}^{f}(\tau)}{\lambda-S^{*}(\tau)}+K\label{eq:34}\end{align}
where $K$ is a large enough constant. Now, define $S^{w}(\tau)$
as \[
S^{w}(\tau)=\arg\max_{\pi\in\crS_{\max}}\sum_{n}\dotprod{W^{f}(\tau)}{\pi}.\]
That is, $S^{w}(\tau)$ is the maximum weight schedule with weight
of queue $n$ as $W_{n}^{f}(\tau)$. Consider the following:\begin{align}
 & \dotprod{W^{f}(\tau)}{\lambda-S^{*}(\tau)}\nonumber \\
= & \dotprod{W^{f}(\tau)}{\lambda-S^{w}(\tau)}+\dotprod{W^{f}(\tau)-U^{f}(\tau)}{S^{w}(\tau)-S^{*}(\tau)}\nonumber \\
 & +\dotprod{U^{f}(\tau)}{S^{w}(\tau)-S^{*}(\tau)}.\label{eq:35}\end{align}
From the definition of $S^{*}(\tau)$, $S^{w}(\tau)$, $\Delta(\tau)(=W(\tau)-U(\tau)),$
and bi-Lipschitz continuity of $f(\cdot)$ it follows that \begin{align}
\dotprod{U^{f}(\tau)}{S^{w}(\tau)-S^{*}(\tau)} & \leq0\label{eq:36}\\
\dotprod{W^{f}(\tau)-U^{f}(\tau)}{S^{w}(\tau)-S^{*}(\tau)} & \leq\rho\dotprod{\Delta(\tau)}{\mathbf{1}}\label{eq:37}\end{align}
Now, for strictly admissible $\lambda$ such that $\lambda=\sum_{k}\alpha_{k}\pi^{k}$
with $\sum_{k}\alpha_{k}=1-\gamma$ for some $\gamma\in\left(0,1\right),$
we obtain that \begin{align}
 & \dotprod{W^{f}(\tau)}{\lambda-S^{w}(\tau)}\nonumber \\
=\; & \dotprod{W^{f}(\tau)}{\sum_{k}\alpha_{k}\pi^{k}}-\left(\gamma+\sum_{k}\alpha_{k}\right)\dotprod{W^{f}(\tau)}{S^{w}(\tau)}\nonumber \\
=\; & \sum_{k}\alpha_{k}\dotprod{W^{f}(\tau)}{\pi^{k}-S^{w}(\tau)}-\gamma\dotprod{W^{f}(\tau)}{S^{w}(\tau)}\label{eq:38}\end{align}
Since $S^{w}(\tau)$ is the maximum weight schedule with weight of
queue $n$ as $W_{n}(\tau)$:\begin{equation}
\dotprod{W^{f}}{\pi^{k}-S^{w}(\tau)}\leq0\quad\forall k\label{eq:39}\end{equation}
Thus, it follows from (\ref{eq:38}) and (\ref{eq:39}) that \begin{equation}
\dotprod{W^{f}(\tau)}{\lambda-S^{w}(\tau)}\leq-\gamma\dotprod{W^{f}(\tau)}{S^{w}(\tau)}.\label{eq:40}\end{equation}
Now since all $N$ entries can be covered by $N$ distinct feasible
schedules, it follows that the weight of maximum weight matching is
at least $1/N$ the sum of weights of all the entries. That is \begin{align}
\dotprod{W^{f}(\tau)}{S^{w}(\tau)} & \geq\frac{1}{N}\sum_{n}W_{n}^{f}(\tau)\nonumber \\
 & =\frac{\abs{W^{f}(\tau)}}{N}\geq\frac{1}{\rho}\frac{\abs{W(\tau)}}{N}\label{eq:41}\end{align}
The last inequality follows from the bi-Lipschitz continuity of $f(\cdot)$.
Combining \eqref{eq:34}-\eqref{eq:41} and taking further expectation
with respect to $W(\tau),$ we obtain \begin{align}
 & \E{\sum_{n}\lambda_{n}F\Bigl(\tilde{W_{n}}(\tau+1)\Bigr)-\sum_{n}\lambda_{n}F\Bigl(W(\tau)\Bigr)}\nonumber \\
\leq\; & \varepsilon\E{\abs{W(\tau)}}+\rho\E{\abs{\Delta(\tau)}}+K,\label{eq:42}\end{align}
where $\varepsilon=\frac{\gamma}{\rho N}.$ To complete the proof,
note that if queue $n$ is non-empty after service at time $\tau$,
then $\tilde{W_{n}}(\tau+1)=W_{n}(\tau+1).$ Else, $W_{n}(\tau+1)=0$
and thus it follows from \eqref{eq:12} that $F\Bigl(\tilde{W_{n}}(\tau+1)\Bigr)\geq0=F(0)=F\left(W_{n}(\tau+1)\right).$
This inequality along with \eqref{eq:42} implies the desired claim
of Lemma~\ref{lem:lyapunov}. 
\end{IEEEproof}

\begin{IEEEproof}
[Proof of Lemma~\ref{lem:WCP}] This result corresponds to a constraint-free
network in which scheduling at different output queues is independent.
Hence, we will prove the result for a single queue operating under
a WCP and strictly admissible loading. We use the same notation, but
with subscript $m$'s dropped. For a single queue operating under
a WCP and strictly admissible loading, busy cycle lengths form an
i.i.d process i.e., $B^{k}$ are i.i.d. We now look at the large deviation
behavior of this process. For a particular $k$ and time $t=0$ starting
from the beginning of busy cycle $B^{k}$, let $I(t)$ denote the
cumulative arrival process during $B^{k}$. Now consider the event
$B^{k}>t$. If the length of the busy cycle is greater than $t$,
it implies that the queue has been non-empty up to time $t$. Further,
since the service process is work conserving, it follows that there
has been one departure every time slot and hence a total of $t$ departures
up to time $t$. Since the total number of departures cannot be more
than the total number of arrivals, it follows that $I(t)>t$. Thus,
we conclude that the event $B^{k}>t$ implies the event $I(t)>t$.
For large enough $t$, we can now write\begin{equation}
\Pr\left(B^{k}>t\right)\leq\Pr\left(I(t)-t>0\right)\leq C\exp\left(-Dt\right)\label{eq:43}\end{equation}
where $C$ and $D$ are some non-negative constants. The last inequality
follows from Chernoff bound, which can be used because arrivals happen
according to a Bernoulli process. Let $\Theta$ denote the random
variable $\max_{k\leq\tau}B^{k}$. Then, we have the following:\begin{align}
\E{\Theta}=\sum_{t}\Pr\left(\Theta>t\right) & =\sum_{t<\Gamma}\Pr\left(\Theta>t\right)+\sum_{t\geq\Gamma}\Pr\left(\Theta>t\right)\nonumber \\
 & \leq\Gamma+\sum_{t\geq\Gamma}\Pr\left(\Theta>t\right)\label{eq:44}\end{align}
 \eqref{eq:44} is true for any non-negative integer $\Gamma$. In
particular, choose $\Gamma$ large enough such that \eqref{eq:43}
is true $\forall t\geq\Gamma$. It now follows from union bound that
\begin{align}
\sum_{t\geq\Gamma}\Pr\left(\Theta>t\right) & \leq\sum_{k\leq\tau}\sum_{t\geq\Gamma}\Pr\left(B^{k}>t\right)\nonumber \\
 & \leq O\left(\tau\exp(-D\Gamma)\right)\label{eq:45}\end{align}
The second inequality follows from \eqref{eq:43}. Now by choosing
$\Gamma=O\left(\log\tau\right)$ we can bound $\sum_{t\geq\Gamma}\Pr\left(\Theta>t\right)$
by $1$. It now follows from \eqref{eq:45} that \[
\E{\max_{k\leq\tau}B^{k}}\leq O\left(\log\tau\right).\]

\end{IEEEproof}

\section{Experiments\label{sec:Experiments}}

We carried out simulations to evaluate the performance of our algorithm
in the context of IQ switches. We assumed a FIFO queuing policy at
the input ports of the IQ switch. We compared the performance of our
algorithm with the Longest Queue First (LQF) and Oldest Cell First
(OCF) algorithms. We used the fixed-length packet switch simulator
available at http://klamath.stanford.edu/tools/SIM/.

We first explain the simulation setting: The switch size is $N=16$.
The buffer sizes are infinite. The policy used is FIFO. All inputs
are equally loaded on a normalized scale, and $\rho\in\left(0,1\right)$
denotes the normalized load. The arrival process is Bernoulli i.i.d.
We use a Uniform load matrix, i.e., $\lambda_{ij}=\rho/N$ $\forall i,j$.
We ran our simulation for 2.1 million time steps removing the first
$100,000$ time steps to achieve steady-state.

Because we are approaching this problem from the perspective of fairness,
we evaluate the aforementioned switching algorithms in terms of Latency
and Output-Queue (OQ) Delay. OQ delay is defined as the difference
of the departure times of a packet in the input queued switch and
the shadow OQ switch. Further, the goal cannot only be to achieve
a better expected latency, but in fact, we wish to value consistency,
or relatively few deviations from the mean. One measure for this are
higher moments of the variables. Thus, here we provide plots for the
logarithm of first and second moments of both Latency and the OQ Delay
versus a uniform load of $\rho$. Figures 5 and 6 plot respectively
the logarithm of the first and second moments of the latency. We observe
that for lower loads, i.e., for $\rho<0.45$ the performance of all
the three algorithms is almost the same. But for higher loads, the
first moment of LQF and MUCF are better than OCF. Fig. 6 shows that
in terms of the second moment, MUCF performs the best and LQF the
worst, with OCF lying between. This is in line with our expectations
because, as mentioned earlier LQF is not fair and hence performs badly
at higher moments. MUCF performs better than OCF for both the moments.

Figures 5 and 6 correspond to latency and figures 7 and 8 correspond
to OQ delay. We observe that MUCF performs better than the other two
algorithms for both the metrics at all the loads, especially for the
second moments illustrating fairness. Thus, the simulations illustrate
that MUCF tracks the performance of an OQ switch better than LQF and
OCF.

\begin{figure}
\includegraphics[width=1\linewidth,height=7cm]{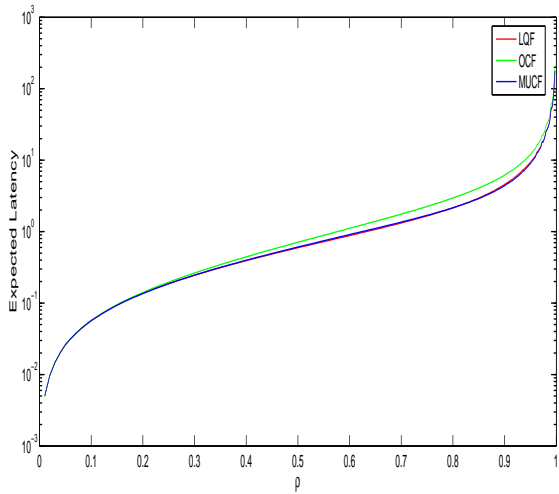}

\caption{Comparison of the logarithm of Expected latencies of different scheduling
algorithms\label{fig:ELA1}}

\end{figure}

\begin{figure}
\includegraphics[width=1\linewidth,height=7cm]{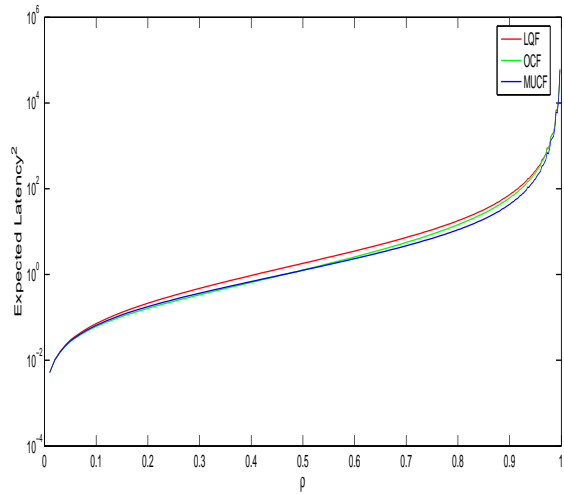}

\caption{Comparison of the logarithm of second moments of the latencies of
different scheduling algorithms. \label{fig:ELA2}}

\end{figure}

\begin{figure}
\includegraphics[width=1\linewidth,height=7cm]{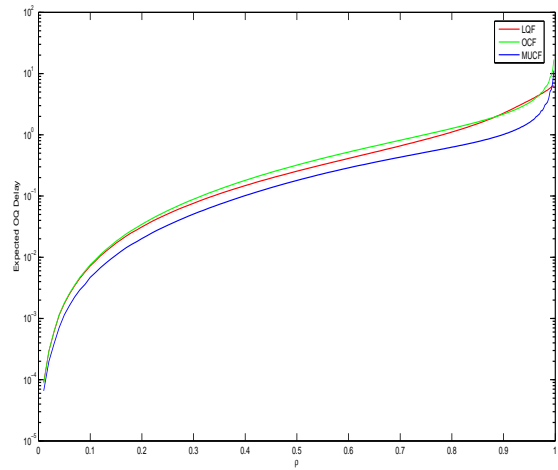}

\caption{Comparison of the logarithm of expected output queued delays of different
scheduling algorithms. \label{fig:EOQ1}}

\end{figure}

\begin{figure}
\includegraphics[width=1\linewidth,height=7cm]{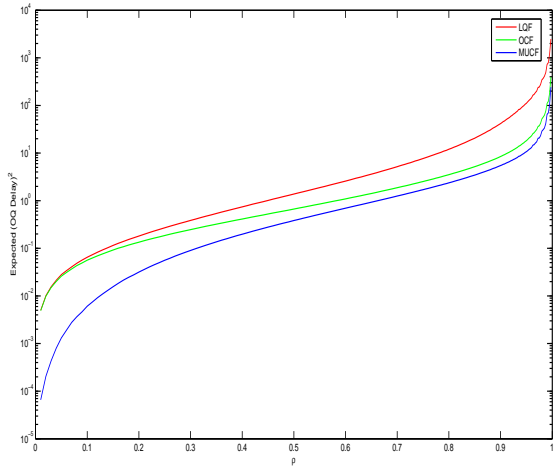}

\caption{Comparison of the logarithm of second moments of the output queued
delays of different scheduling algorithms. \label{fig:EOQ2}}

\end{figure}

\section{Conclusion\label{sec:Conclusion}}

In this paper, we considered the problem of designing a fair scheduling
algorithm for constrained queuing systems. Fairness in networks is
not only an intuitively desired goal, but also one with many practical
benefits. Most of the existing work concentrates on fairly allocating
bandwidth to different flows in the network. A major limitation of
this approach is that it disregards the packetized nature of flows.
We overcame this problem and proposed a packet based notion of fairness
by establishing a novel analogy with the ranked election problem.
Ranked election is a widely studied problem in the Economics literature,
and this analogy allowed us to leverage that work. This results in
a packet based notion of fairness and an algorithm to achieve this
fairness.

Rather surprisingly, the algorithm turned out be the familiar MW style
algorithm. Moreover, it does not require the knowledge of flow arrival
rates. Our fairness algorithm also fits into the utility maximization
framework that is more popular for designing fair algorithms. This,
in some sense, validates our approach. We also proved that our algorithm
is throughput optimal. This result is very crucial since the emulation
approach already achieves fairness, but with a loss of throughput.
Also, the proof is non-trivial and requires some non-traditional techniques
to be introduced because existing proof techniques don't directly
apply. We believe that the proof techniques we introduced are more
widely applicable to similar problems and this is another important
contribution of the paper. Finally, our simulation results corroborate
the fact that our algorithm is better at providing fairness than the
more popular algorithms in the context of input queued switches.

\bibliographystyle{ieeetr}

\begin{IEEEbiographynophoto}{Srikanth Jagabathula} 
Srikanth Jagabathula received the BTech degree in Electrical Engineering from the Indian Institute of Technology (IIT) Bombay in 2006, and the MS degree in Electrical Engineering and Computer Science from the Massachusetts Institute of Technology (MIT), Cambridge, MA in 2008. He is currently a doctoral student in the Department of Electrical Engineering and Computer Science at MIT. His research interests are in the areas of revenue managemenr, choice modeling, queuing systems, and compressed sensing. He received the President of India Gold Medal from IIT Bombay in 2006. He was also awarded the "Best Student Paper Award" at NIPS 2008 conference and the Ernst Guillemin award for the best EE SM Thesis. \end{IEEEbiographynophoto}

\begin{IEEEbiographynophoto}{Devavrat Shah} Devavrat Shah is currently a Jamieson career development associate professor with the department of electrical engineering and computer science, MIT. He is a member of the Laboratory of Information and Decision Systems (LIDS) and affiliated with the Operations Research Center (ORC). His research focus is on theory of large complex networks which includes network algorithms, stochastic networks, network information theory and large scale statistical inference.
He received his BTech degree in Computer Science and Engg. from IIT-Bombay in 1999 with the honor of the President of India Gold Medal. He received his Ph.D. from the Computer Science department, Stanford University in October 2004. He was a post-doc in the Statistics department at Stanford in 2004-05. 
He was co-awarded the best paper awards at the IEEE INFOCOM '04, ACM SIGMETRICS/Performance '06; and best student paper awards at Neural Information Processing Systems '08 and ACM SIGMETRICS/Performance '09. He received 2005 George B. Dantzig best disseration award from the INFORMS. He received the first ACM SIGMETRICS Rising Star Award 2008 for his work on network scheduling algorithms.  \end{IEEEbiographynophoto}
\end{document}